\documentclass[3p
 ]
{elsarticle}
\usepackage{lineno,hyperref}
\modulolinenumbers[5]
\usepackage[cp1251]{inputenc}
\biboptions{sort&compress}
\usepackage{graphicx}
\usepackage{bm}
\usepackage[T1]{fontenc}
\usepackage{amsmath}
\def\be{\begin{equation}}
\def\ee{\end{equation}}
\def\nn{\nonumber}
\def\d{{\rm d}}

\def\prl{\parallel}

\begin{document}
\begin{frontmatter}
\title{The angular momentum of electron radiation in a uniform magnetic field}
\author{Vladimir Epp\corref{mycorrespondingauthor}}
\cortext[mycorrespondingauthor]{Corresponding author}
\ead{epp@tspu.edu.ru}
\author{Ulyana Guselnikova}
\ead{guselnikova.ulyana@yandex.ru}

\address{Tomsk State Pedagogical University, ul. Kievskaya 60, 634061 Tomsk, Russia}

\begin{abstract}
We study theoretically by means of quantum electrodynamics the vortex radiation of a relativistic electron in a uniform magnetic field.
The exact expressions for the probability of emission of a photon with a certain angular momentum are found. The classical asymptotics  $\hbar\to 0$ of this probability does not match the angular momentum flux density calculated by the classical method using the symmetrized energy-momentum tensor.
Although the flux of angular momentum integrated over the radiation directions is the same in both cases. We found the angular momentum flux of the radiation field using the canonical (not symmetrized) energy-momentum tensor and showed that the flux obtained in this way coincide with the classical limit  for the probability of photon emission. 
\end{abstract}

\begin{keyword}
\texttt{vortex radiation\sep angular momentum\sep electron \sep synchrotron radiation\sep Dirac equation}
\MSC[2010] 78A40\sep 	33E20  
\end{keyword}

\end{frontmatter}


\section*{Introduction}
\addcontentsline{toc}{section}{Introduction}
Electromagnetic radiation carrying an angular momentum of more than one Planck constant per photon attracts considerable attention of researchers due to the possibility of application in various fields of physics: information transfer, interaction with atoms, high-energy particle collisions and radiation processes. Vortex light beams have opened up a wide range of applications such as spatial optical confinement of atoms or microscopic objects and phase contrast microscopy \cite{Barnett2017,Andrews2012book}. The first theoretical and experimental studies of twisted light or vortex radiation were devoted to laser radiation modified by an astigmatic optical system or numerically calculated holograms \cite{Allen1992, Padgett2004}.
These publications stimulated extensive research of the vortex optical beams. A more extensive bibliography on the history in this area can be found in the reviews \cite{Barnett2017, Hernandez-Garcia2017}.

One of the most promising sources of vortex radiation in the X-ray and the gamma range of the spectrum  are the charged particles moving in a spiral or vortex beams of charged particles \cite{Karlovets_2015,Silenko_2017}. In particular, spiral motion is realized in a uniform magnetic field and in a helical undulator. Various schemes for obtaining a beam of twisted photons in undulators \cite{Sasaki2008, Bordovitsyn2012, Matsuba_2018} and free electron lasers \cite{Hemsing2009,Hemsing2011,Hemsing2012} have been proposed. Analytical calculations have shown that a particle moving in a spiral emits  photons  carrying an average angular momentum equal to $n\hbar$ with $n$ being the harmonic number \cite{Katoh2017, EppGuselnikova2019, Salehi_2021}.
 These theoretical results have been confirmed by a number of experimental studies of radiation in helical undulators \cite{Bahrdt2013, Kaneyasu2017, Katoh_2017-SR}.

In the works listed above  the angular momentum of the radiation of a particle moving along a spiral was studied within the framework of classical electrodynamics. The quantum aspects of this problem were studied in \cite{Bogdanov2018, Bogdanov2019PhysRev} by use of  semiclassical approach.

In the present work, we calculated the probability of radiation of photons with a high angular momentum using the exact solution of the Dirac equation for an electron in a uniform magnetic field. This article is a continuation of our work \cite{EppGuselnikova2019}, where the angular momentum of radiation at spiral motion  was studied using classical methods.

The article is organized as follows. The section \ref{one} contains the well-known solution of the Dirac equation for an electron in a uniform magnetic field. The main purpose of this section is to remind the wave function of the electron and to introduce the notation that will be used in what follows. In the  section \ref{two}, we calculate the probability of spontaneous transitions with the emission of a photon in a given direction, carrying an orbital angular momentum. The classical limit ($\hbar\to 0$) is discussed in section \ref{three}.

The result obtained here differs from the angular momentum flux density calculated earlier by the classical method. Therefore, in section  \ref{four}, we find the canonical angular momentum flux density in the Coulomb gauge of the vector potential by use of the classical method and show that this angular momentum flux is consistent with the quantum  calculation. Finally, in the last section, we discuss the obtained results.

\section{Solution to the Dirac equation}\label{one}
 Let an electron with charge $e=-e_0$ and mass $m$ be in a uniform magnetic field $\bm{H}$. In a cylindrical coordinate system $r, \varphi, z$ with the $z$ axis directed along the magnetic  field, the vector potential can be given as
 \[
 \bm A=\frac 12 Hr\bm{\hat\varphi},
 \]
 where $\bm{\hat\varphi}$ is the unit vector corresponding to the $\varphi$ coordinate. The solution to the Dirac equation in such a field has the form \cite{Rabi_1928, Ternov_1966, SokolovTernov}
 \begin{align}\label{psi}
\Psi_{nl\zeta k_z}(\bm r,t)=\sqrt{\frac{\gamma}{\pi L}}\exp[i(-Kct+(l-1/2)\varphi+k_z z)]
\left(\begin{array}{r}
c_1e^{-i\varphi /2}I_{n-1,s}(\rho)\\
c_2e^{i\varphi /2}I_{n,s}(\rho)\\
c_3e^{-i\varphi /2}I_{n-1,s}(\rho)\\
c_4e^{i\varphi /2}I_{n,s}(\rho)\end{array}\right),
\end{align}
where $n$ and $l$ are the principal and orbital quantum numbers, respectively, $n=0,1,2..$, $s$ is the radial quantum number: $s=n-l$. Spin number $\zeta=1$ if the electron spin is directed along the magnetic field and $\zeta=-1$ -- vice versa. The wave function (\ref{psi}) is an eigenfunction of the
$z$-component of the  angular momentum operator $\widehat {L}_z$  and of the $z$-component of the linear momentum operator $\widehat {p}_z$ 
\[
\widehat {L}_z\Psi=\left(l-\frac{i}{2}\right)\hbar\Psi, \quad\hat{p}_z  \Psi=k_z\Psi,
 \]
$\hbar$ is the Planck constant. The following notations are used in (\ref{psi}) 
\[
\gamma=\frac{e_0H}{2c\hbar}, \quad K=\frac{E}{c\hbar},\quad \rho=\gamma r^2,
\]
$c$ is the speed of light, $E$ is the particle energy 
\be\label{E}
E=c\sqrt{m^2c^2+\hbar^2(k_z^2+4\gamma n)},
\ee
$L$ is the formal length restricting motion along the $z$ axis, $I_{ns}(\rho)$ are the Laguerre functions
\be\label{I}
I_{ns}(\rho)=\sqrt \frac{s!}{n!}e^{-\rho/2}\rho^{(n-s)/2}Q_s^{n-s}(\rho),
\ee
and $Q_s^l$ are the generalized Laguerre polynomials
\[
Q_s^l=\frac{1}{s!}e^\rho \rho^{-l}\frac{\d^s}{\d\rho^s}(\rho^{s+l}e^{-\rho}).
\]
The $c_i$ matrix determines the electron spin polarization
\begin{equation*}
\begin{pmatrix}
c_1 \\
c_2 \\
c_3 \\
c_4
\end{pmatrix}
=\frac{1}{2\sqrt{2}}
\begin{pmatrix}
B_+(A_++A_-) \\
B_-(A_--A_+) \\
B_+(A_+-A_-) \\
B_-(A_++A_-)
\end{pmatrix}
,
\end{equation*}
where
\begin{align}
A_+=&\sqrt{1+k_z/K}\, , \quad A_-=\zeta\sqrt{1-k_z/K}\, ,\\
B_+=&\sqrt{1+\zeta k_0/K_0}\,,\quad B_-=\zeta\sqrt{1-\zeta k_0/K_0}\,,
\end{align}
and
\[
K_0=\sqrt{K^2-k_z^2}\,, \quad k_0=mc/\hbar.
\]
\section{Probability of the transitions $l\to l'$}\label{two}
Let us find the probability of emission of a photon with a wave vector $\bm {k}$ and a $z$-component of angular momentum  equal to $L_z$. By virtue of the angular momentum conservation law, this is the probability of transition from the state with the quantum number $l$ to the state with the quantum number $l'=l-L_z/\hbar$. We will specify the photon polarization by the basis
\be\label{e}
\bm e_\pm=\frac{1}{\sqrt{2}}(\bm e_1\pm i\bm e_2),
\ee
where $\bm e_1$ and $\bm e_2$ are mutually orthogonal real vectors orthogonal to the wave vector $\bm{k}$, so that the vectors listed in the order $\bm e_1, \bm e_2, \bm{k }$  follow the right-hand rule. The vector $\bm e_+$ corresponds to the right-handed polarization,  and $\bm e_-$ to the left-handed one.

The average value of  $z$-component of the  photon angular momentum  is found by 
\be\label{LL}
\frac{dL_z}{\d t}=\sum_{m'}\hbar(l-l')w_\pm,
\ee
where $m'=n',l',k'_z, \zeta'$ are the quantum numbers of the final state, and $w_\pm$ is the probability of  transitions from state $|nlk_z\zeta\rangle$  to state $|n'l'k'_z\zeta'\rangle$.
The transition probability $w_\pm$ was calculated in \cite{Ternov_1966, SokolovTernov, Bordovitsyn-SR}. Here we briefly repeat the main steps of the calculations.
\be\label{w1}
w_\pm=\frac{e_0^2}{2\pi \hbar}\int\frac{\d^3k}{k}\delta(K-K'-k)S_\pm.
\ee
The matrix elements $S_\pm$ in the basis (\ref{e}) take the form 
\be\label{S}
S_\pm=\frac{1}{2}[((\bm{n}\times\langle\bm{\alpha}\rangle^*)(\bm{n}\times\langle\bm{\alpha}\rangle))\pm((\bm{n}(\langle\bm{\alpha}\rangle\times\langle\bm{\alpha}\rangle^*))],
\ee
The asterisk denotes complex conjugate. The vector $\langle\bm{\alpha}\rangle$ represents the matrix elements of the Dirac matrix $\bm{\alpha}$
\be\label{alpha}
\langle\bm{\alpha}\rangle=\int\psi_{n'l'k'_z\zeta}l^{-(\bm{k}\bm{r})}\bm{\alpha}\Psi_{nl\zeta k_z}\d^3x,
\ee
The vector $\bm{n}=\bm{k}/k$ is the unit vector in the direction of photon linear momentum. By virtue of axial symmetry, the vector $\bm{n}$ can be specified as $(n_x, n_y,n_z)=(0, \sin\theta, \cos\theta)$ where $\theta$ is the angle between the $ z$-axis and vector $\bm{n}$.
Integration in (\ref{w1}) over $d^3k$ leads to the expression
\be\label{w2}
w_\pm=\frac{e_0^2}{\hbar}\int S_\pm\frac{k \sin\theta d \theta}{1-\beta_{||}\cos\theta},
\ee
where
\[
\beta_{\prl}=-\frac{\partial K'}{\partial k}=\frac{k_z-k\cos\theta}{K-k} .
\]
In the spherical coordinate system $r,\theta,\varphi$ the matrix elements (\ref{S}) are
\begin{align}\label{Spm}
S_\pm=&\frac{1}{2}\left[|\langle\alpha_1\rangle|^2+|\langle\alpha_2\rangle\cos\theta-\langle\alpha_3\rangle\sin\theta|^2\mp iS_\prl\right],\\
S_\prl=&\bigl(\langle\alpha_1\rangle^*\langle\alpha_2\rangle-\langle\alpha_2\rangle^*\langle\alpha_1\rangle\bigr)
\cos\theta\nn\\
&-\bigl(\langle\alpha_1\rangle^*\langle\alpha_3\rangle-\langle\alpha_3\rangle^*\langle\alpha_1\rangle\bigr)\sin\theta\nn.
\end{align}
Integration in  (\ref{alpha}) over the volume gives
\begin{align}
\langle\alpha_1\rangle&=i[(c'_1c_4+c'_3c_2)I_{n,n'-1} -(c_1c'_4+c_3c'_2)I_{n-1,n'} ]f,&\nn\\
 \langle\alpha_1\rangle&=[(c'_1c_4+c'_3c_2)I_{n,n'-1} -(c_1c'_4+c_3c'_2)I_{n-1,n'} ]f,&\nn\\
\langle\alpha_3\rangle&=[(c'_1c_3+c_1c'_3)I_{n-1,n'-1} -(c_2c'_4+c'_2c_4)I_{n,n'} ]f,&\label{mat}
\end{align}
where the functions $I_{n,n'}=I_{n,n'}(x)$ are the functions of the argument $x$
\be\label{xq}
x=\frac{k^2\sin^2\theta}{4\gamma}, \quad f=I_{ss'}(x)\delta_{k'_z, k_z-k\cos\theta}.
\ee
Finally, substituting (\ref{w2}) into (\ref{LL}), we obtain the rate of radiation  of the $z$-component of angular momentum
\be\label{Lz}
\frac{\d L_z}{\d t}=e_0^2\sum_{m'}(l-l')\int S_\pm \frac{k\sin\theta \d\theta}{1-\beta_{||}\cos\theta}.
\ee
This expression differs from the formula for the intensity of radiation  \cite{Bordovitsyn-SR,SokolovTernov}
\[
W=ce_0^2\sum_{m'}\int S_\pm\frac{k^2\sin\theta\d \theta}{1-\beta_\prl\cos\theta}
\]
by replacing the $z$-component of the  photon angular momentum $\hbar(l-l')$ by the photon energy $\hbar\omega$.

The modulus of the wave vector $k$ in  (\ref{Lz}) is found as a solution to the system of equations expressing the energy and momentums  conservation law
\begin{align}\label{k_z}
&k_z-k_z'-k\cos\theta=0,\\
&K-K'-k=0.
\end{align}
This gives
\begin{align}\label{k-1}
k=K\frac{1-\beta_\prl\cos\theta}{\sin^2\theta}\left(1-\sqrt{1-\frac{4\gamma(n-n')\sin^2\theta}{K^2(1-\beta_\prl\cos\theta)^2}}\,\right).
\end{align}
Equations (\ref{Spm}) -- (\ref{Lz}) determine the angular momentum flux carried by photons during spontaneous  transitions to lower energy levels.
\section{The classic limit}\label{three}
Next, we find the asymptotic expression for the angular momentum flux (\ref{Lz}) in the limit  $\hbar\to0$ and $n, n'\gg 1$. In this approximation, the wave vector (\ref{k-1}) is equal to
\begin{align}\label{k-2}
k\approx\frac{(n-n')\omega_0}{c(1-\beta_\prl\cos\theta)},
\end{align}
where $\omega_0=eHc/E$ is the frequency of electron motion in a circle. It can be seen from  (\ref{E}) that the linear momentum component orthogonal to the $z$ axis is equal to $p_\perp=2\hbar\displaystyle\sqrt{\gamma n}$. In the classical limit $p_\perp=Ev_\perp/c^2$, where $v_\perp$ is the velocity part orthogonal to the $z$ axis. Hence, it is convenient to introduce the notation
\begin{align}\label{b}
\beta_\perp=\frac{v_\perp}{c}=\frac{2c\hbar\sqrt{\gamma n}}{E}
\end{align}
Substituting (\ref{k-2}) and (\ref{b}) into the  argument of the Laguerre functions (\ref{xq}), we obtain
\be\label{x}
x\approx \frac{(n-n')^2\beta_\perp\sin^2\theta}{4n(1-\beta_\prl\cos\theta)}.
\ee
In the classical limit, the photon energy is much less than the electron energy, i.e. $n-n' \ll n$. Therefore, for $n\to\infty$ and fixed $n-n'$ the variable $x$ decreases as $1/n$. The asymptotic expression for Laguerre polynomials of the form $Q_n^l(z/n)$ for large $n$ and fixed $l$ and $z$ has the form
  \cite{Sokolov_1955, Szego_1975}
\be
Q_n^l\left(\frac{z}{n}\right)\approx n^l e^{z/2n}z^{-l/2}J_l(2\sqrt{z}),
\ee
where $J_l$ are the Bessel functions. Hence,
\begin{align}\label{I}
I_{n,n'}\approx J_{n-n'}(\xi),\quad \xi=\frac{\nu\beta_\perp\sin\theta}{1-\beta_\prl\cos\theta}.
\end{align}
This asymptotics does not apply to the Laguerre functions $I_{s,s'}(x)$, since in the latter case $x$ tends to zero regardless of the indices $s$ and $s'$. Assuming $s$ and $s'$ to be fixed and letting $x$ tend to zero, we obtain for the Laguerre polynomials
\begin{align}\label{l-Q}
\lim_{x\to 0}Q_n^l (x)=\frac{1}{n!}(l+1)(l+2)\dots (l+n).
\end{align}
Substituting this into the Laguerre function, we get
\begin{align}\label{l-I}
\lim_{x\to 0}I_{s,s'}(x)=\delta_{s,s'}.
\end{align}
Thus, we can the neglect the change in quantum number $s$ in transitions between states with large  numbers $n$. In this case $n-n'=l-l'$.
This means that the $z$-component of the photon angular momentum can only be positive (if $z$-axis is directed along the magnetic field).
Substituting the resulting expansions into  (\ref{Spm}), averaging the coefficients $c_i$ over the initial  spin states, and summing up $c_i'$ over the final states, we obtain
\begin{align}\label{s}
S_\pm=\frac 12\left[\beta_\perp J'_\nu(\xi)\pm\frac{\cos\theta-\beta_\prl}{\sin\theta} J_\nu(\xi)\right]^2.
\end{align}
As a result, the flux of angular momentum in the radiation field  (\ref{Lz}) takes the form
\be\label{dL-1}
\frac{\d L_{z\pm}}{\d t}=\frac{e_0^2\omega_0}{2 c}\sum_{\nu=1}^\infty \nu^2\int\limits_0^\pi\left[\beta_\perp J'_\nu(\xi) 
\pm\frac{\cos\theta-\beta_\prl}{\sin\theta} J_\nu(\xi)\right]^2
\frac{\sin\theta\d\theta}{(1-\beta_\prl\cos\theta)^2}.
\ee
This expression was obtained in  \cite{Bogdanov2018} by use of semiclassical method.

The integrand in  Eq. (\ref{dL-1}) differs from the flux density of  angular momentum obtained by use of classical electrodynamics \cite{EppGuselnikova2019}, which reads (for simplicity, we put $\beta_\prl=0$):
\be\label{sch11}
\frac{\d L_{z}}{\d\Omega\d t}=\frac{e_0^2\omega_0\beta\sin \theta}{2\pi c}\sum_{\nu=1}^\infty \nu\left\{ \xi\left[{J'_\nu}^2(\xi)+\cot^2\theta\, J^2_\nu(\xi)\right]
+J_\nu(\xi) J'_\nu(\xi)\right\}.
\ee
The main difference between the Eqs (\ref{dL-1}) and (\ref{sch11}) is that the angular momentum flux (\ref{sch11}) in the direction of the $z$ axis is equal to zero, while the flux (\ref{dL-1}) is not.
According to standard electrodynamics,  the flux of angular momentum component $(\bm L\bm n)$ in the direction of the vector $\bm n$ is equal to zero \cite{Jackson_Cl_El, Sokolov1991}
The angular momentum flux (\ref{sch11}) satisfies this condition. But the angular momentum flux of radiation in the direction of the $z$ axis, obtained by  quantum theory, is not equal to zero. In particular, if we put $\theta=0,\pi$ in Eq. (\ref{dL-1}), we get
\begin{align}\label{dL-0}
\frac{\d L_{z\pm}}{\d t}=\frac{e_0^2\omega_0\beta_\perp}{8\pi c(1-\beta_\prl)^2}\left[1 \pm \frac{\cos\theta-\beta_\prl}{1-\beta_\prl\cos\theta} \right]_{\theta=0,\pi}.
\end{align}
This means that  in the direction of the $z$ axis only right-handed photons are emitted, and only left-handed ones are emitted in the opposite direction.

More detailed information can be obtained from the exact quantum equations (\ref{Spm}) -- (\ref{Lz}). For $\theta=0,\pi$ the argument of the Laguerre functions (\ref{xq}) vanishes and $I_{ss'}(0)=\delta_{ss'}$. Provided that $n'<n$, the matrix elements (\ref{mat}) take the form
\begin{align}\label{alp}
\langle \alpha_1\rangle&=-i(c_1c'_4+c_3c'_2)\delta_{n-1,n'}\delta_{s,s'}\delta_{k'_z,k_z-k\cos\theta},\\
\langle \alpha_2\rangle&=i \langle \alpha_1\rangle, \quad \langle \alpha_3\rangle=0.
\end{align}
As we can see, the matrix elements are nonzero only for $n' = n-1$ and $s'=s$ and, consequently, for $l'=l-1$. Thus, in the $\theta=0,\pi$ directions, photons are emitted only at transitions $|n,l,\zeta\rangle\to|n-1,l-1,\zeta'\rangle$. Decrease the quantum number $l$ by one means that the angular momentum of these photons  is $\hbar$, and the projection on the direction of radiation is either $\hbar$ at $\theta=0$, or $-\hbar$ at $\theta=\pi$.
Hence, the angular momentum of photons emitted in the direction of the $z$ axis is represented only by  spin. Their orbital momentum is zero. Which, in fact, is consistent with the classical representation (\ref{dL-1}), but not consistent with (\ref{sch11}). Note again that the results of integration over the angles of the Eqs (\ref{dL-1}) and (\ref{sch11}) are the same. The reasons for this discrepancy are discussed in the next section.

\section{Canonical angular momentum}\label{four}
The disagreement  between the quantum and classical results obtained in the previous section is due to the fact that the angular momentum density in classical electrodynamics is defined ambiguously. The straightforward application of
the Noether theorem to electromagnetic field yields the proper integral conservation laws.
 In this case, the integrands can be transformed in such a way that the integral over the entire space does not change. In particular, the symmetrization of the energy-momentum tensor of the electromagnetic field  is based on this \cite{Belifante_1939, Landau_II}.

When calculating the angular momentum flux in our paper \cite{EppGuselnikova2019}, we proceeded from the symmetrized energy-momentum tensor, which leads to the well-known expression for the angular momentum flux in the wave zone  \cite{Jackson_Cl_El, Panofsky1962}
\be\label{L}
\frac{\d\bm L}{\d t}=\frac{r^2}{c}\int(\bm r\times \bm P)\d\Omega,
\ee
where
\[
\bm P=\frac{c}{4\pi}(\bm E\times \bm H)
\]
is the Poynting vector, and $\bm E$ and $\bm H$ are the electric and magnetic fields respectively.

Meanwhile, an immediate consequence of Noether's theorem is the conservation of the nonsymmetric canonical energy-momentum tensor \cite{Jackson_Cl_El, Landau_II}
\[
T^{\mu\nu}=-\frac{1}{4\pi}g^{\mu\alpha} F_{\alpha\rho}\partial^\nu A^\rho+\frac{1}{16\pi}g^{\mu\nu}F_{\rho\sigma}F^{\rho\sigma}
\]
The corresponding density of the canonical angular momentum is represented by the  third rank tensor \cite{Bjorken_1965} 
\[
M^{\alpha\beta\gamma}= \left(x^\alpha T^{\beta\gamma}-x^\beta T^{\alpha\gamma}\right)+\frac{1}{4\pi }\left(-A^\alpha F^{\beta\gamma}+A^\beta F^{\alpha\gamma}\right).
\]
In this case, the angular momentum density of the electromagnetic field is given by the component $M^{0\beta\gamma}$ or by the vector $M_i=\varepsilon_{ijk}M^{0jk}$, where $ j,k=1,2,3$, and $\varepsilon_{ijk}$ is the unit antisymmetric tensor of the third rank. Accordingly, the angular momentum flux density in the wave zone is equal to $c \bm M$, and the total flux is determined by the integral
\be\label{Lcan}
\frac{\d \bm { L}}{\d t}=cr^2\int\bm M\d\Omega.
\ee
The vector $\bm M$ in the Coulomb gauge $\bm \nabla\bm A=0$ can be represented as the sum \cite{Jackson_Cl_El, Bliokh_2013, Afanasev_2022}
\be\label{Mvec}
\bm M=\bm{\mathcal L}+\bm{\mathcal S},
\ee
where $\bm{\mathcal L}$ and $\bm{\mathcal S}$  are the ``orbital'' and ``spin'' parts respectively
 \be\label{MM}
 \bm {\mathcal L}=\frac{1}{4\pi c}E_i(\bm r\times\bm\nabla)A_i,\quad \bm {\mathcal S}= \frac{1}{4\pi c}(\bm E\times\bm A). 
 \ee

The result of integration over the solid angle in Eqs (\ref{L}) and (\ref{Lcan}) is the same, however the integrands differ significantly.
Indeed, let us calculate the angular momentum flux due to (\ref{Lcan}), using the expressions for the Fourier components of the electric field and vector potential obtained in \cite{EppGuselnikova2019}
 \begin{align}\label{An1r}
A_{\nu r}&=\frac{B}{r}J_\nu(\xi), \quad  A_{\nu\theta}=\frac{B}{ r}\cot\theta J_\nu(\xi), \nn\\
  A_{\nu\varphi}&=i\frac{B\beta}{r} J'_\nu(\xi), \quad B=e e^{i (k r+\nu\varphi-\nu\pi/2)}\\
 E_{\nu r}&=0,\quad  E_{\nu\theta}=\frac{i B\nu\omega\cot\theta}{rc} J_\nu(\xi),\quad E_{\nu\varphi}=-\frac{ B\nu\omega\beta}{ rc}J'_\nu(\xi).\nn
 \end{align}
 When using the vector potential, it suffices to restrict ourselves to quantities of the first order in powers of $1/r$. In the Coulomb gauge one should put $A_{\nu r}=0$.
Let us substitute these values into (\ref{MM}). In order to find the time-averaged values for the $\nu$-th harmonic, we need to make the substitution $\bm E\to \bm E^*_\nu,\,\bm A\to\bm A_\nu$. As expected, the vector $\bm{\mathcal L}_\nu$ has only a transverse component ($\bm{\mathcal L}_\nu\cdot\bm r=0$), while the vector $\bm{ \mathcal S}_\nu$ is directed radially ($\bm{ \hat r}=\bm r/r$):
\[
\bm{\mathcal S}_{\nu}=\frac{e_0^2\nu\omega_0\beta}{2\pi r^2 c^2}\cot\theta J'_\nu(\xi)J_\nu(\xi)\bm{ \hat r}. 
\]
The $z$-component of the angular momentum flux is equal to
\begin{align}\label{mm}
{\mathcal L}_{\nu z}&=\frac{e_0^2\nu^2\omega_0}{ 4\pi r^2 c^2}\left[\beta^2 {J'}^2_\nu(\xi) +\cot^2\theta J^2_\nu(\xi)-\frac{2\beta}{\nu}\frac{\cos^2\theta}{\sin\theta}J'_\nu(\xi)J_\nu(\xi)\right],\\
{\mathcal S}_{\nu z}&=\frac{e_0^2\nu\omega_0\beta}{2\pi r^2 c^2}\frac{\cos^2\theta}{\sin\theta}J'_\nu(\xi)J_\nu(\xi)
\end{align}
Substituting these expressions into (\ref{Lcan}) and (\ref{Mvec}), we obtain the total flux of the $z$-component of angular momentum
\begin{align}\label{dL-2}
\frac{\d L_{z}}{\d t}=\frac{e_0^2\omega_0}{2 c}\sum_{\nu=1}^\infty \nu^2\int\limits_0^\pi\left[\beta^2 {J'}^2_\nu(\xi) +\cot^2\theta J^2_\nu(\xi)\right]
\sin\theta\d\theta
\end{align}
in full accordance with Eq. (\ref{dL-1}) obtained from the probability of photon emission in the limit  $\hbar\to 0, \,\beta_\prl=0$.

In the direction $\theta=0$, the $z$-component of the ``orbital'' angular momentum  is equal to zero, and in the $z$-component of the ``spin'' angular momentum, only the first harmonic remains
\[
\bm{\mathcal S}_{\nu}\Big|_{ \theta=0}=\frac{e_0^2\omega_0\beta^2}{8\pi r^2 c^2}\delta_{1\nu}\bm{ \hat r}.
\]
Wherein, the average angular momentum per one photon is equal to $\hbar$. In the direction $\theta=\pi/2$, which lies in the plane of the electron orbit, on the contrary, the ``spin'' angular momentum ${\mathcal S}_{\nu z}$ is equal to zero, and the ``orbital'' angular momentum ${\mathcal L}_{\nu z}$ takes on a maximum value. The absence of ``spin'' angular momentum is due to the fact that in the classical limit the probability of emission of left-handed photons is equal to the probability of emission of right-handed ones in this direction. In this case, the ``orbital'' angular momentum is associated with the momentum transfer of the electromagnetic field.

\section{Conclusion}
Using the solution to the Dirac equation for an electron in a uniform magnetic field, we have found exact expressions for the probability of emission of a photon with a certain projection of its angular momentum onto the direction of the magnetic field. We have also found the asymptotic expression for the angular momentum flux density in  radiation field in the limit $\hbar\to 0$. The expressions obtained in this way do not coincide with the previously found equations for the angular momentum flux density derived from the symmetrized energy-momentum tensor of the electromagnetic field.
This discrepancy is explained by the fact that the density of the angular momentum of the field and, accordingly, the flux of the angular momentum are determined ambiguously. For comparison, we found the angular momentum flux in radiation using the canonical (not symmetrized) energy-momentum tensor and showed that the formulas obtained in this way coincide with the classical limit of equations for the probability of photon emission. In this case, the flux of angular momentum integrated over the radiation directions following from the canonical and symmetrized energy-momentum tensor, as expected, coincide.

The question of the extent to which the canonical or symmetrized energy-momentum tensor is applicable to real physical measurements is traditionally the subject of hot discussions \cite{Teitelboim1980, Bak_1994, Ohanian_1986,Bordovitsyn2012, Bliokh_2013,Obukhov_2022-Sy}. This problem is closely
related to the determination of the angular momentum density of the field and to the separation of spin and orbital degrees of freedom.
Obviously, the applicability of certain formulas to specific physical measurements is determined by the specific conditions of the light–matter interactions.  
 Currently, various methods for registering the orbital angular momentum of radiation are being developed \cite{Chen2015, Li2019,Wakamatsu_2021}. Moreover, there are indications of the possibility of independent measurement of the spin and orbital angular momentum \cite{VanEnk_1994, ONeil2002,Allen_1999}. 

\section*{Acknowledgments}
This research was supported  by Russian Foundation for Basic Research, project No. 19-42-700011. 

%

\end{document}